\documentclass[11pt,envcountsame]{llncs}
\usepackage{cite}
\usepackage{amsmath}
\usepackage{amssymb}
\usepackage{mathrsfs}
\usepackage{tikz}
\usepackage[all]{xy}
\newcommand{\inv}{^{-1}}
\title{The averaging trick and the \v{C}ern\'y conjecture}
\author{Benjamin Steinberg\thanks{The author gratefully acknowledges the support of NSERC}
\institute{School of Mathematics and Statistics\\ Carleton University\\ 1125 Colonel By Drive\\ Ottawa, ON, Canada \\ \email{bsteinbg@math.carleton.ca}}}
\date{May 8, 2010}
\spnewtheorem{Thm}{Theorem}{\bfseries}{\itshape}
\spnewtheorem{Lemma}[Thm]{Lemma}{\bfseries}{\itshape}
\spnewtheorem{Conjecture}[Thm]{Conjecture}{\bfseries}{\itshape}
\spnewtheorem{Rmk}[Thm]{Remark}{\itshape}{\rmfamily}
\begin{document}
\maketitle

\begin{abstract}
The results of several papers concerning the \v{C}ern\'y conjecture are deduced as consequences of a simple idea that I call the averaging trick.  This idea is implicitly used in the literature, but no attempt was made to formalize the proof scheme axiomatically.  Instead, authors axiomatized classes of automata to which it applies.
\end{abstract}

\section{Introduction}

Recall that a (complete deterministic) automaton $\mathscr A=(Q,\Sigma)$ with state set $Q$ and alphabet $\Sigma$ is called \emph{synchronizing} if there is a word $w\in \Sigma^*$ such that $|Qw|=1$.  The word $w$ is called a \emph{synchronizing word}.  The main conjecture in this area is:

\begin{Conjecture}[\v{C}ern\'y~\cite{cerny}]
An $n$-state synchronizing automaton admits a synchronizing word of length at most $(n-1)^2$.
\end{Conjecture}

There is a vast literature on this subject.  See for example~\cite{Pincerny,pincernyconjecture,synchgroups,dubuc,cerny,volkovc1,rystsov1,rystsov2,AMSV,trahtman,traht2,volkovc2,Kari,volkovc3,rystcom,rystrank,mycerny,Karicounter,VolkovLata,PerrinBeal,strongtrans,strongtrans2,mortality,beal,Salomcerny}. The best known upper bound is cubic~\cite{twocomb}, whereas it is known that one cannot do better than $(n-1)^2$~\cite{cerny}.

My goal here in this note is not to prove the \v{C}ern\'y conjecture for a new class of automata, but rather to give a no-frills, uniform approach to an argument that underlies a growing number of results in the \v{C}ern\'y conjecture literature (cf.~\cite{rystsov1,Kari,beal,PerrinBeal,strongtrans,strongtrans2}).  Underlying all these results (as well as the more difficult results of~\cite{dubuc} and~\cite{mycerny}) are two simple ideas:
\begin{itemize}
\item if a finite sequence of numbers is not constant, then it must at some place exceed its average;
\item finite dimensional vector spaces satisfy the ascending chain condition on subspaces.
\end{itemize}
The latter idea is often cloaked in the language of rational power series.

The paper is organized as follows.  In the next section I state what I call the ``Averaging Lemma.''  It is a method, with a probabilistic flavor, for obtaining bounds on lengths of synchronizing words.  Before proving the lemma, I show how to deduce from it Kari's solution of the \v{C}ern\'y conjecture for Eulerian automata, as well as recent results of B\'eal and Perrin~\cite{PerrinBeal} for one-cluster automata and Carpi and d'Alessandro~\cite{strongtrans,strongtrans2} for (locally) strongly transitive automata.  We also recover an old result of Rystsov~\cite{rystsov1} on regular automata (which is essentially the same thing as strongly transitive automata). In fact, we obtain new generalizations of all these results. The final section proves the Averaging Lemma.

\section{The averaging trick}
Let $\Sigma$ be an alphabet.  Denote by $\Sigma^*$ the free monoid on $\Sigma$ and put \[\Sigma^{\leq d} = \bigcup_{m=0}^d \Sigma^m.\] The ring of polynomials with real coefficients in the non-commuting variables $\Sigma$ is denoted $\mathbb R\Sigma$.  By a (finitely supported) \emph{probability} on $\Sigma^*$, we mean an element \[P=\sum_{w\in \Sigma^*}P(w)w\in \mathbb R\Sigma\] such that: $P(w)\geq 0$ for all $w\in \Sigma^*$, and
\[\sum_{w\in \Sigma^*}P(w)=1.\]
The \emph{support} of $P$ is \[\sigma(P)=\{w\in \Sigma^*\mid P(w)>0\}.\]
Notice that if $P_1$ and $P_2$ are probabilities, then so is $P_1P_2$.  Also note that $\sigma(P_1P_2)=\sigma(P_1)\sigma(P_2)$.

If $X\colon \Sigma^*\to \mathbb R$ is a \emph{random variable}, then the \emph{expected value} of $X$ (with respect to the probability $P$) is:
\begin{equation}\label{expectedval}
\mathbf E_P(X) = \sum_{w\in \Sigma^*}P(w)X(w) = \sum_{w\in \sigma(P)}P(w)X(w).
\end{equation}
The fundamental property of a random variable that we exploit in this paper is that either it is almost surely constant (and equal to its expectation), or with positive probability it exceeds it expectation.  More precisely,
it is immediate from \eqref{expectedval} and the definition of a probability that either $X(w)= \mathbf E_P(X)$ for all $w\in \sigma(P)$, or there is a value $w\in \sigma(P)$ with $X(w)>\mathbf E_P(X)$.

Suppose now that $\mathscr A=(Q,\Sigma)$ is an automaton with $|Q|=n$.  We view elements of $\mathbb RQ$ as row vectors.  Let $\pi\colon \mathbb R\Sigma\to M_n(\mathbb R)$ be the corresponding matrix representation (cf.~\cite{BerstelReutenauer}); so if \[f=\sum_{w\in \Sigma^*}f(w)w,\] and $q,r\in Q$, then \[\pi(f)_{q,r} = \sum_{\{w\in \Sigma^*\mid qw=r\}}f(w).\]
We shall usually omit $\pi$ from the notation and view $\mathbb R\Sigma$ as acting on row and column vectors.  If $S\subseteq Q$, then $[S]$ denotes the characteristic row vector of $S$; e.g., $[Q]$ is the all ones row vector.  We use $[S]^T$ to denote the transpose vector.  A key fact is that $w[S]^T=[Sw\inv]^T$ for $w\in \Sigma^*$, where as usual $Sw\inv = \{q\in Q\mid qw\in S\}$.

\begin{Lemma}[Averaging Lemma]\label{expand}
Let $\mathscr A=(Q,\Sigma)$ be a synchronizing automaton with $n$ states, let $P_1$ be a probability on $\Sigma^*$ and let $R\subseteq Q$.  Set $c=2$ if, for each proper non-empty subset $S\subsetneq R$, there exist $w_1,w_2\in \sigma(P_1)$ with $Sw_1\inv\neq Sw_2\inv$ and otherwise put $c=1$. Suppose that there exists a probability $P_2$ with support $\Sigma^{\leq n-c}$ such that:
\begin{enumerate}
\item  $[R]P_2P_1=[R]$;
\item  $R\subseteq q\Sigma^*$ for all $q\in R$;
\item  there exists $w_0\in \Sigma^*$ with $Qw_0\subseteq R$.
\end{enumerate}
Then $\mathscr A$ has a synchronizing word of length at most:
\begin{itemize}
\item $c+(n-2)(n-c+L)$ if $R=Q$;
\item $(r-1)(n-c+L)+\ell+c-1$ if $R\subsetneq Q$
\end{itemize}
where $r=|R|$, $L$ is the maximum length of a word in $\sigma(P_1)$ and $\ell=|w_0|$.
\end{Lemma}

\begin{Rmk}
If $r$ is odd, then the proof shows that the bounds in Lemma~\ref{expand} can be improved to $1+(n-2)(n-c+L)$ and $(r-1)(n-c+L)+\ell$, respectively.
\end{Rmk}

Before, proving the lemma, let us use it to derive anew some results from the literature.  The first is a result of Kari on synchronizing Eulerian automata~\cite{Kari}.  An automaton is \emph{Eulerian} if its underlying graph admits an Eulerian directed path, or equivalently, it is strongly connected and the in-degree of every vertex is the same as the out-degree (and hence is the alphabet size).  Actually, we can generalize his result.

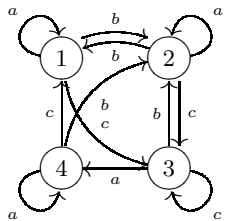
\begin{figure}[hbtp!]
\[\entrymodifiers={++[o][F-]}
\SelectTips{cm}{}
\xymatrix @-1pc
 {1\ar@(l,u)[]^{a} \ar@<1ex>@/^/[rr]^{b}\ar@/_1pc/[rrdd]^c&*{}&2\ar@(r,u)[]_{a}\ar@/_/[ll]^b\ar@<1ex>[dd]^c\\
 *{}&*{}&*{}\\
 4\ar@(l,d)[]_{a} \ar[uu]^{c}\ar@/^1pc/[rruu]_b&*{}&3\ar@(r,d)[]^{c}\ar[ll]^a\ar[uu]^b}\]
\caption{A pseudo-Eulerian automaton\label{pseudoeulerian}}
\end{figure}

Let us say that a strongly connected automaton $\mathscr A=(Q,\Sigma)$ is \emph{pseudo-Eulerian} if we can find a probability $P$ with support $\Sigma$ such that the matrix $\pi(P)$ is doubly stochastic (i.e., each row and column of $P$ adds up to $1$).  For instance, if $\mathscr A$ is Eulerian with adjacency matrix $A$ and $d=|\Sigma|$, then we can set \[P=\sum_{a\in \Sigma}d^{-1}a.\]  One checks that $\pi(P) = d^{-1}A$, and hence is doubly stochastic by the Eulerian hypothesis.  Thus every Eulerian automaton is pseudo-Eulerian.  It is easy to check whether a strongly connected automaton is pseudo-Eulerian:  one just needs to look for a strictly positive solution to the system of $|Q|+1$ linear equations
\begin{alignat*}2
1 &= \sum_{a\in \Sigma}p_a&\\
1 &= \sum_{a\in \Sigma}p_a\cdot |qa\inv|&\quad (q\in Q).
\end{alignat*}
The automaton in Figure~\ref{pseudoeulerian} is pseudo-Eu\-ler\-i\-an but not Eu\-ler\-i\-an.
Indeed, if we put $P=a/2+b/6+c/3$, then \[\pi(P) =\begin{bmatrix}\frac{1}{2} &\frac{1}{6} &\frac{1}{3} &0\\\frac{1}{6} &\frac{1}{2} &\frac{1}{3}  &0\\0&\frac{1}{6}  &\frac{1}{3} &\frac{1}{2}\\ \frac{1}{3}&\frac{1}{6}&0&\frac{1}{2}\end{bmatrix}\] is doubly stochastic.

\begin{Thm}
An $n$-state synchronizing pseudo-Eulerian automaton has a synchronizing word of length at most $1+(n-2)(n-1)$.
\end{Thm}
\begin{proof}
Let $\mathscr A=(Q,\Sigma)$ and suppose that $P$ is a probability with support $\Sigma$ such that $\pi(P)$ is doubly stochastic.   Let $P_1$ be the probability with support concentrated on the empty word and take $R=Q$.  As pseudo-Eulerian automata are strongly connected, $Q\subseteq q\Sigma^*$ for all $q\in Q$. Put \[P_2 = \frac{1}{n}\sum_{m=0}^{n-1}P^m;\] it is a probability with support $\Sigma^{\leq n-1}$.   The condition that $\pi(P)$ is doubly stochastic is equivalent to $[Q]P=[Q]$.  Thus \[[Q]P_2P_1=[Q]\cdot \frac{1}{n}\sum_{m=0}^{n-1}P^m =[Q].\]
The Averaging Lemma now yields the upper bound of $1+(n-2)(n-1)$ on the length of a synchronizing word.
\qed\end{proof}

The next result simultaneously generalizes results of Rystsov~\cite{rystsov1} on regular automata, B\'eal~\cite{beal} on circular automata, B\'eal, Berlinkov and Perrin~\cite{PerrinBeal,bealperrinnew} on one-cluster automata and Carpi and d'Alessandro~\cite{strongtrans,strongtrans2} on strongly and locally strongly transitive automata.

\begin{Thm}\label{rystsovgeneralized}
Let $\mathscr A=(Q,\Sigma)$ be a synchronizing automaton.  Suppose there is a set of words $W\subseteq \Sigma^*$ and $k\geq 1$ so that, for each state $q\in Q$ and each state $s\in R=QW$, there are exactly $k$ elements of $W$ taking $q$ to $s$. Let $\ell$ be the length of the shortest word in $W$ and $L$ be the length of the longest.  If $R=Q$, then there is a synchronizing word for $\mathscr A$ of length at most $2+(n-2)(n-2+L)$; if $R\subsetneq Q$, then there is a synchronizing word of length at most $(r-1)(n-2+L)+\ell+1$ where $r=|R|$.
\end{Thm}
\begin{proof}
A straightforward counting argument establishes that $|W|=kr$. It remains to define our probabilities in order to apply the Averaging Lemma.
Take $P_1$ to be the uniform distribution on $W$ (so $P_1(w)=1/|W|$ for $w\in W$ and is otherwise $0$). To verify that $c=2$, let $\emptyset\neq S\subsetneq R$ and suppose that $s\in S$ and $q\in R\setminus S$.    Then by the hypothesis on $W$, there exist $w_1,w_2\in W$ with $rw_1=s$ and $qw_2=q$.  Then $q\in Sw_1\inv$ but $q\notin Sw_2\inv$.

Now let $P_2$ be an arbitrary probability with support $\Sigma^{\leq n-c}$. The only condition remaining to check in order to apply the Averaging Lemma is that $[R]P_2P_1=[R]$.  First observe that the columns of $\pi(P_1)$ corresponding to elements of $Q\setminus R$ are zero, while if $s\in R$, then the corresponding column of $\pi(P_1)$ is $(k/|W|)[Q]^T=(1/r)[Q]^T$.  Since $\pi(P_2)$ is a stochastic matrix (each of its rows sum to $1$), this means that $\pi(P_2P_1)=\pi(P_1)$.
Next observe that if $s\in R$, then $s\sum_{w\in W}w= k[R]$.  Thus \[[R]\sum_{w\in W}w = \sum_{s\in R}s\sum_{w\in W}w=rk[R]=|W|[R].\]  Therefore, $[R]P_1=[R]$ and hence $[R]P_2P_1=[R]$, as required.
\qed\end{proof}

\begin{figure}
	\centering
	\newlength{\nodedist}
	\setlength{\nodedist}{1.5cm}
	
	\begin{tikzpicture}[shorten >=1pt,node distance=\nodedist,auto]
		\tikzstyle{state}=[circle, draw, fill=black!50, inner sep=0pt, minimum width=4pt]
		
		\foreach \X in {0,...,4}
			\node at (\X*72:\nodedist*0.85065)	[state]		(A\X)		{};

		\node	[state]		(B0)		[right of=A0]			{};
		\node	[state]		(B1)		[below right of=B0]		{};
		\node	[state]		(B2)		[below right of=B1]		{};
		\node	[state]		(B3)		[right of=B1]			{};
		\node	[state]		(B4)		[above right of=B0]		{};
		\node	[state]		(B5)		[right of=B4]			{};
		\node	[state]		(B6)		[above right of=B4]		{};
		\node	[state]		(C0)		[above of=A1]			{};
		\node	[state]		(C1)		[above left of=A2]		{};
		\node	[state]		(C2)		[above left of=C1]		{};

		\path [->]
			(A0)		edge		node	[swap]	{$a$}		(A1)
			(A1)		edge		node	[swap]	{$a$}		(A2)
			(A2)		edge		node	[swap]	{$a$}		(A3)
			(A3)		edge		node	[swap]	{$a$}		(A4)
			(A4)		edge		node	[swap]	{$a$}		(A0)
			(B0)		edge		node	[swap]	{$a$}		(A0)
			(B1)		edge		node	[swap]	{$a$}		(B0)
			(B2)		edge		node	[swap]	{$a$}		(B1)
			(B3)		edge		node	[swap]	{$a$}		(B1)
			(B4)		edge		node	[swap]	{$a$}		(B0)
			(B5)		edge		node	[swap]	{$a$}		(B4)
			(B6)		edge		node	[swap]	{$a$}		(B4)
			(C0)		edge		node	[swap]	{$a$}		(A1)
			(C1)		edge		node	[swap]	{$a$}		(A2)
			(C2)		edge		node	[swap]	{$a$}		(C1)
			;

	\end{tikzpicture}
	\caption{$a$-skeleton of a one-cluster automaton with $n=15$ and $r=5$.}
	\label{oneclustfig}
\end{figure}
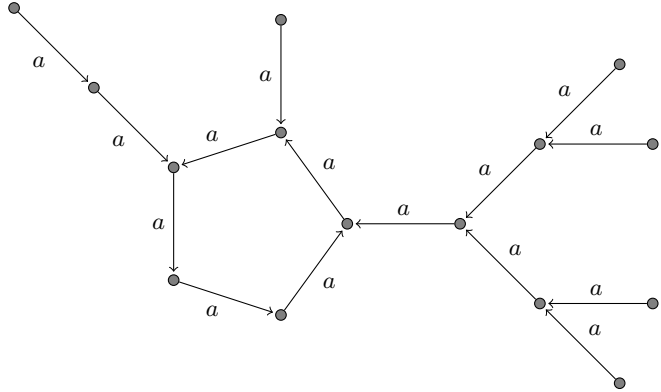

For example, B\'eal and Perrin~\cite{PerrinBeal} call $\mathscr A=(Q,\Sigma)$ a \emph{one-cluster automaton} if there exists $a\in \Sigma$ so that $a$ has only one cycle $R$ on $Q$; see Figure~\ref{oneclustfig}.  Suppose that the cycle has size $r$.  Then each state of $Q$ is taken to exactly one element of $R$ by the set of words $W=\{a^{n-r},\ldots,a^{n-1}\}$.
Theorem~\ref{rystsovgeneralized} then yields the bound of $2n^2- 7n+8$.  This should be compared with the bound of $2n^2-7n+7$ from~\cite{bealperrinnew}, which improves on the earlier bound of $2n^2-6n+5$ from~\cite{PerrinBeal}. Indeed, if $r=n$, Theorem~\ref{rystsovgeneralized} immediately yields a bound of $2+(n-2)(2n-3)=2n^2-7n+8$.  Otherwise, using $L=n-1$ and $\ell=n-r$, we obtain a bound of
\begin{align*}
(r-1)(2n-3)+n-r+1
&=r(2n-4)-n+4\\ &\leq (n-1)(2n-4)-n+4\\ &=2n^2-7n+8.
\end{align*}
Similarly, one recovers the results of Rystsov~\cite{rystsov1} and the results of Carpi and d'Alessandro~\cite{strongtrans,strongtrans2} with an improved bound.  Indeed, the locally strongly transitive automata of~\cite{strongtrans2} constitute the special case of Theorem~\ref{rystsovgeneralized} where $k=1$.  Rystsov's notion of a regular automaton is essentially (but slightly more rigid) than the case $R=Q$.

The proof of Theorem~\ref{rystsovgeneralized} can easily be adapted to obtain the same bound if $W$ is an arbitrary set of words such that there is a probability $P_1$ supported on $W$ so that each column of $\pi(P_1)$ corresponding to an element of $Q\setminus R$ is $0$, whereas each column corresponding to an element of $R$ is $1/r[Q]^T$.

\section{Proof of the Averaging Lemma}
The proof of the Averaging Lemma rests on our observation about expectations of random variables
and the ascending chain condition for finite dimensional vector spaces.  Suppose that $\Sigma^*$ acts on the left of a vector space $V$ by linear maps.  Let $X\subseteq \Sigma^*$ and let $W$ be a subspace.  Then by $XW$, we mean the span of all vectors $xw$ with $x\in X$ and $w\in W$.

\begin{Lemma}\label{ascending}
Let $\pi\colon \Sigma^*\to M_n(K)$ be a matrix representation with $K$ a field.  Suppose that one has subspaces $W,V\subseteq K^n$ of column vectors with $W\subseteq V$, but $\Sigma^*W\nsubseteq V$.  Let $S$ be a spanning set for $W$.  Then there exist $s\in S$ and $w\in \Sigma^*$ with $|w|\leq \dim V-\dim W+1$ and $ws\notin V$.
\end{Lemma}
\begin{proof}
Put $W_m = \Sigma^{\leq m}W$.  Then there is an ascending chain of subspaces \[W=W_0\subseteq W_1\subseteq W_2\subseteq \cdots\] and moreover as soon as this chain stabilizes it equals $\Sigma^*W$.  By our assumption, there is a greatest $m\geq 0$ with $W_m\subseteq V$.  In particular, the chain does not stabilize until after $m$ steps and so \[W_0\subsetneq W_1\subsetneq \cdots \subsetneq W_m\subseteq V\] and hence $\dim W_0+m\leq \dim V$, that is, $m+1\leq \dim V-\dim W+1$.  Therefore, there is a word $w\in \Sigma^*$ with $|w|\leq \dim V-\dim W+1$ and $wW\nsubseteq V$.  But $W$ is spanned by $S$, so we can find $s\in S$ with $ws\notin V$.
\end{proof}

\begin{proof}[of the Averaging Lemma]
For convenience, put $X=\sigma(P_1)$. We show that for each $\emptyset\neq S\subsetneq R$, there exists $w\in \Sigma^*$ of length at most $n-c+L$ with $|Sw\inv\cap R|>|S|$ except for when $c=2$ and $|S|=r/2$, in which case we can only guarantee that $w$ has length at most $n-1+L$.  If $R=Q$, the result is then immediate:  one can find a state $q\in Q$ and a letter $a\in \Sigma$ so that $|qa\inv|>1$;  now we expand by inverse images $n-2$ times with words of length at most $n-c+L$ (except for when $c=2$ and $|S|=r/2$, in which case we expand by $n-1+L$) to obtain the result.  If $R\subsetneq Q$, we can find $w$ of length at most $(r-1)(n-c+L)+c-1$ with $|Rw|=1$ using the same idea. Then as $Qw_0\subseteq R$, it follows $|Qw_0w|\leq |Rw|=1$.  This yields the bound of $(r-1)(n-c+L)+\ell+c-1$ on the length a synchronizing word.

Consider the probability $P=P_2P_1$ on $\Sigma^*$ and define a random variable $Z_S\colon \Sigma^*\to \mathbb R$ by \[Z_S(w) =|Sw\inv \cap R|=[R][Sw\inv]^T=[R][w][S]^T.\]  Let us compute the expected value of this random variable:
\begin{align*}
\mathbf E_P(Z_S) = &\sum_{w\in \Sigma^*}P(w)|Sw\inv \cap R|= \sum_{w\in \Sigma^*}P(w)[R]w[S]^T
\\ &= [R]P[S]^T = [R]P_2P_1[S]^T=[R][S]^T
\\ &=|S|
\end{align*}
where we have used $[R]P_2P_1=[R]$.  The support of $P$ is $\sigma(P_2)\sigma(P_1) = \Sigma^{\leq n-c}X$.  If we can find $v\in \Sigma^{\leq n-c}X$ with $Z_S(v)=|Sv\inv \cap R|\neq |S|$, then we can find $w\in \Sigma^{\leq n-c}X$ with $|Sw\inv \cap R|=Z_S(w)>|S|$ by our discussion earlier on random variables that are not almost surely constant.  As $|w|\leq n-c+L$, this will finish the proof.

If $|Sx\inv\cap R|\neq |S|$ for some $x\in X$, then we are done.  Otherwise, we may assume $|Sx\inv\cap R|=|S|$ for all $x\in X$.  Let $\gamma$ be the column vector $[S]^T-(|S|/r)[Q]^T$.  Notice that if $w\in \Sigma^*$, then one has \mbox{$w\gamma = [Sw\inv]^T-(|S|/r)[Q]^T$} and so $[R]w\gamma= |Sw\inv\cap R|- |S|$.  In particular, if $x\in X$ our assumption implies $[R]x\gamma=0$.  Moreover, $x\gamma\neq 0$ as $|S|<r$.   Thus if $W$ is the subspace spanned by the column vectors $x\gamma$ with $x\in X$, then $0\neq W\subseteq [R]^{\perp}$.

Our next goal is to verify that $\dim W\geq c$ unless $c=2$ and $|S|=r/2$ (in which case it is at least $1$).  The only non-trivial case is when $c=2$ and $|S|\neq r/2$.  Then we can find $w_1,w_2\in X$ with $Sw_1\inv\neq Sw_2\inv$.  We claim that $w_1\gamma$ and $w_2\gamma$ are linearly independent elements of $W$.  Indeed, if they were linearly dependent, then since both vectors are non-zero we must have $w_1\gamma=kw_2\gamma$ for some $k\in \mathbb R$.  Moreover, $k\neq 1$ because $Sw_1\inv\neq Sw_2\inv$.  Thus $[Sw_1\inv]^T-k[Sw_2\inv]^T=(|S|/r)(1-k)[Q]^T$.  Since $[Q]^T$ is the all ones column vector and $[Sw_1\inv]^T$, $[Sw_2\inv]^T$ are column vectors of zeroes and ones, it follows that $k=-1$ and $Sw_1\inv, Sw_2\inv$ are complementary subsets of $Q$.  Then we obtain $[Q]^T=(2|S|/r)[Q]^T$, whence $|S|=r/2$, a contradiction.
We conclude that $w_1\gamma$ and $w_2\gamma$ are linearly independent and so  $\dim W\geq 2=c$.

Our next claim is that $\Sigma^*W\nsubseteq [R]^{\perp}$.  Indeed, let $w$ be a synchronizing word.  Then $ww_0$ synchronizes $\mathscr A$ to an element of $q\in R$.  But $q\Sigma^*\supseteq R$, so we can synchronize to any state of $R$.  In particular, we can synchronize $\mathscr A$ via some word $y$ into $Sx\inv\cap R$ for some $x\in X$.  Then $Sx\inv y\inv = Q$ and so $[R]yx\gamma=|Sx\inv y\inv\cap R|-|S|>0$.   This shows that $yx\gamma\notin R^{\perp}$ and hence $\Sigma^*W\nsubseteq [R]^{\perp}$.
As $\dim W\geq c$ and $\dim [R]^{\perp}=n-1$, Lemma~\ref{ascending} now provides $u\in \Sigma^{\leq n-c}$ and $z\in X$ with $uz\gamma\notin [R]^{\perp}$.  Putting $v=uz\in \Sigma^{\leq n-c}X$, we have $0\neq [R]v\gamma =|Sv\inv \cap R|-|S|$.  This completes the proof.
\qed\end{proof}

\begin{Rmk}
The above proof and the proof of the main result of~\cite{bealperrinnew} give an improved bound for one-cluster automata.  It is shown in~\cite{bealperrinnew} that if we have an $n$-state one-cluster automaton with unique $a$-cycle $R$ of length $r$, then we can find a state $q\in R$ and a word $w$ of length at most $2n-r-1$ such that $|qw\inv\cap R|>1$.  Since the \v{C}ern\'y conjecture is proved for the case $r=n$~\cite{dubuc}, we may assume $r\leq n-1$.  Combining this with the above proof yields a bound of \begin{align*}(r-2)(2n-3)+2n-r-1+n-r+1 &=(r-2)(2n-3)+3n-2r\\ &=r(2n-5)-n+6\\ &\leq (n-1)(2n-5)-n+6\\&= 2n^2-8n+11.\end{align*}
\end{Rmk}

\section*{Acknowledgments}
I would like to thank Volker Diekert for a suggestion that led to the improvement of the bound in the Averaging Lemma in the case $c=2$, as well as for detecting a minor flaw in an earlier version of this paper.  Steffen Kopeck produced the diagram in Figure~\ref{oneclustfig}.


\end{document}